\documentstyle[psfig,graphics]{mn2e}
\title [Kinematics of the chromospherically active binaries]
{Kinematics of the chromospherically active binaries
and evidence of an orbital period decrease in binary evolution}

\author[Karata\c{s} et al.]
       {Y.~Karata\c{s},$^1 \thanks{E-mail: karatas@istanbul.edu.tr}$
        S.~Bilir$^1$, Z. Eker$^{2}$\thanks{Visiting Astronomer, 
        Istanbul University Science Faculty, Department of Astronomy and Space
        Sciences} and O.~Demircan$^{3}$\footnotemark[2]\\
  $^1$Istanbul University Science Faculty, Department of Astronomy and Space
      Sciences, 34452 University-Istanbul, Turkey\\
  $^2$Department of Physics and Astronomy, King Saud University, PO Box 2455, 
     Riyadh, Saudi Arabia\\
  $^3$\c{C}anakkale Onsekiz Mart University Observatory, 17100 \c{C}anakkale, 
      Turkey\\}

\date{Accepted 2003 12 29.
      Received 2004 04 02;
      }

\pagerange{\pageref{firstpage}--\pageref{lastpage}}
\begin{document}

\maketitle

\label{firstpage}

\begin{abstract}

Kinematics of 237 Chromospherically Active Binaries (CAB) were 
studied. The sample is heterogeneous with different orbits and 
physically different components from F to M spectral type main 
sequence stars to G and K giants and super giants. The computed 
$U$, $V$, $W$ space velocities indicate the sample is also heterogeneous 
in the velocity space. That is, both kinematically younger and 
older systems exist among the non-evolved main sequence and the 
evolved binaries containing giants and sub giants. The 
kinematically young (0.95 Gyr) sub-sample (N=95), which is formed 
according to the kinematical criteria of moving groups, was 
compared to the rest (N=142) of the sample (3.86 Gyr) in order 
to investigate observational clues of the binary evolution.
Comparing the orbital period histograms between the younger and 
older sub-samples, evidences were found supporting Demircan's 
(1999) finding that the CAB binaries lose mass (and angular 
momentum) and evolve towards shorter orbital periods. The evidence 
of mass loss is noticeable on the histograms of the total mass 
($M_{h}+M_{c}$), which is compared between the younger (available 
only N=53 systems) and older sub-samples (available only N=66 
systems). The orbital period decrease during binary evolution is 
found to be clearly indicated by the kinematical ages of 6.69, 5.19, 
and 3.02 Gyr which were found in the sub samples according to 
the period ranges of $logP\leq0.8$, $0.8<logP\leq1.7$, and 
$1.7<logP\leq3$ among the binaries in the older sub sample. 

\end{abstract}


\section{Introduction}

Chromospherically Active Binaries (CAB) are the 
class of detached binary systems with spectral 
types later than F characterized by a strong chromospheric, 
transition region, and coronal activity. Enhanced emission 
cores of Ca II H and K, and sometimes in the Balmer $H_{\alpha}$
line are primary indicators of the chromospheric activity and 
often accompanied by photometric variability due to starspots. 
The first published catalog of CAB by Strassmeier et al. (1988) 
contained the classical RS CVn systems defined by Hall (1976) and
the BY Draconis-type binaries defined by Bopp \& Fekel (1977). 
The 168 CAB in the first catalog have been increased to 206 in 
the second catalog by Strassmeier et al. (1993). The third catalog
has not yet been published but Eker privately continued to collect
CAB. The unpublished list of Eker, which became the main database 
for this study, nowadays contains about 280 CAB.

At the first attempt, Eker (1992), was rather unsuccessful in 
breaking  up the 146 CAB sample into kinematically distinct sub 
samples. Containing spectral types from F to M and luminosity 
classes from V to II, the already heterogeneous CAB sample was 
found to be heterogeneous also in the sense that the kinematically 
younger and older systems exist among the evolved binaries with 
at least one component of a giant or a sub giant, and among the 
un-evolved main sequence binaries. The Hipparcos data 
(Perryman et al. 1997) was not available to Eker (1992). 
However, Aslan et al. (1999), who used the Hipparcos 
proper motions and parallaxes of 178 CAB, also found 
no clues to non homogeneity in the velocity dispersions. 
Aslan et al. (1999) concluded that there are not any 
significant differences between the sub samples of 
RS CVn binaries, although there are some indications 
of the main sequence RS CVn binaries having smaller 
velocity dispersions, indicating the smaller ages.

The increased number of the CAB sample with the greatly 
improved astrometric data (parallaxes and proper motions) 
of Hipparcos (Perryman et al.\ 1997) motivated us to restudy
the kinematics of the CAB in a similar manner to Eker (1992). 
As soon as  the $U$, $V$, $W$ space velocities and the dispersions 
were produced, the ($\gamma$) shaped concentration in the 
velocity space near the location of the local standard of
rest (LSR) on the $(U, V)$ plane was immediately noticed. 
It was soon realized that the concentration was formed by
young binaries belonging to the moving groups.

The Moving Groups (MG) are kinematically coherent groups 
of stars that share a common origin, and thus offer a better 
way of compiling sub samples of CAB with the same age. 
Eggen (1994) defined a supercluster as a group of stars
gravitationally unbound but sharing the same kinematics 
occupying extended regions in the Galaxy. Therefore, a moving group,
unlike the well known open clusters, can be observed all over 
the sky. Determination of the possible members of MG among the 
binaries and single stars was carried out by 
Eggen (1958a-b, 1989, 1995), Montes et al.\ 
(2001a) and King et al.\ (2003).

Consequently, the initial intention of studying 
the kinematics of the CAB sub samples like Eker (1992) 
was changed to break up the whole sample into two groups,
where the first group contains the possible MG members 
chosen by the kinematical criteria originally defined by 
Eggen (1958a, b, 1989, 1995) and the rest of the sample.
Picking up the possible MG members from the whole sample 
of CAB which are known to be young made it possible to form
kinematically young and old sub samples. After studying the
kinematics and determining the average ages, the histograms
of the total mass ($M_{h}+M_{c}$), period, mass ratio and 
orbital eccentricity were compared between the two sub samples 
as much as the available data permits. This new system of 
investigation using kinematical data made it possible to 
discover new observational clues to the binary evolution 
confirming that the detached CAB also lose mass and angular 
momentum. The angular momentum loss and period decrease were 
predicted for the tidally locked short period systems 
(Demircan 1999). Apparently, the binary evolution
with orbital angular momentum loss also 
exists among the unlocked long period systems. Due to 
the limited space, the investigation into the mass loss 
rates and associated rates of orbital period decrease 
will be handled in a forthcoming study. 

\section{Data}
The 237 systems were selected out of the 280 CAB in 
the unpublished list of Eker. The criteria of selection
is the possession of complete  basic data, 
(proper motion, parallax, and radial velocity) 
allowing computation of the space velocity of 
a binary system with respect to the Sun. 
The selected systems are listed in Table 1 with the 
columns indicating an order number, the most common name, 
HD and Hipparcos cross reference numbers, celestial coordinates 
(ICRS J2000.0), proper motion components, parallax and radial
velocity. The basic data were displayed with associated
standard errors. The reference numbers in the last column 
are separated into three fields with semicolons to indicate
from where the basic data were taken. The two or more 
reference numbers in a field separated by commas indicate 
sub fields if there is more than one reference to any basic data.

\subsection {Parallaxes and proper motions}
The parallaxes and the proper motions in Table 1 were 
taken mainly from {\it The Hipparcos and Tycho Catalogs\/} (ESA 1997)
and {\it The Tycho Reference Catalog\/} (Hog et al.\ 1998). Among 
the 237 systems in Table 1, only 15 (6.3\%) binaries do not have 
Hipparcos parallaxes. Most of the Hipparcos parallaxes have 
relative errors much less than 50\% ($\sigma_{\pi}/\pi<<
0.5$). Only 14 systems (5.9\%) in our list have $\sigma_{\pi}/\pi>0.5$. 
Care was taken not to use parallaxes 
less than the two-sigma detection limit of Hipparcos which
is 1.94 mas ($\sigma$=0.97 mas) (Perryman et al. 1997). 
It was therefore decided to discard the parallax measurement 
of five systems (IN Com, HD122767, RT CrB, V832 Her, AT Cap)
and treat them as the other 15 binaries without a trigonometric 
parallax. However, the other nine systems with $\sigma_{\pi}/\pi>0.5$ 
(V764 Cen, RV Lib, HD152178, V965 Sco, CG Cyg, RS Umi, RU Cnc,
SS Cam, V1260 Ori) were kept in the main list since their 
parallaxes are bigger than the detection limit.

For the systems without trigonometric parallaxes, a published parallax  
of any kind was preferred. 
Among the 20 (15 with no $\pi$, five below the detection limit) only the 
six systems (HP Aur, HZ Com, HD71028, HD122767, V846 Her and V1430 Aql)
were found without any published parallax so that the spectroscopic 
parallaxes were estimated for them from their spectral types and 
luminosity classes.

The Hipparcos and the Tycho catalogues usually 
supply an associated uncertainty for all of the measurements 
of the parallax and the proper motion components. However, 
there are six systems in the list without an uncertainty at 
the proper motion components. `No errors quoted' may imply something 
odd about the star. One possibility is that no errors were there 
because none could be established. On the other hand, it could be a 
simple omission or too few data to permit a standard error estimation. 
With an optimistic approach, we have preferred to adopt the announced 
average uncertainties, which are 0.88 mas/yr in 
$\mu_{\alpha} cos\delta$ and 0.74 mas/yr in $\mu_{\delta}$, 
by Perryman et al. (1997) for the Hipparcos stars brighter 
than ninth magnitude. However, the uncertainty of 5.5 mas/yr 
in the proper motion components for HP Aur were taken from 
Nesterov et al. (1995). Similarly, 2.5 mas/yr uncertainty is 
taken from Bakos et al. (2002) for the systems $\xi$ Uma B and CM Dra.

Nevertheless, the major contribution into the propagated errors  
for the U,V,W space velocities comes from the uncertainty of the parallax. 
Therefore, the largest errors must be associated with the nine systems 
(3.8\% in the list) with $\sigma_{\pi}/\pi>0.5$. In order to see their effect, 
an average propagated uncertainty of those nine systems were computed as 
$\delta U=\pm 7.16$, $\delta V=\pm 11.09$ and $\delta W=\pm 6.94$ km/s. 
However, there is a large intrinsic spread in the galactic space motions (U,V,W) 
that even such large uncertainties emposed by several individual motions appear 
to be unimportant. But still, for the sake of the statistical completeness, 
the missing standard errors of 15 spectroscopic parallaxes had to be completed. 

Sparke \& Gallagher (2000) state that if the interstellar absorption and the reddening 
do not introduce problems, the luminosities of the main-sequence stars can 
often be found to within 10\%, leading to 5\% uncertainties in their distance. 
The giant branch is almost vertical, thus the best hope for determining a luminosity 
is within 0.5 in the absolute magnitude, and hence the distance to 25\%. 
Being in the safe side, the sub giants were assumed to be as giants, thus 25\% 
uncertainty were assigned for the missing standard errors of eight giants and four sub 
giants. The missing standard errors of three systems (IM Vir, HP Aur, HZ Com) with dwarf 
components were assigned with a 5\% uncertainty as Sparke \& Gallagher (2000) suggest. 
With a median distance of 98 pc, the current CAB sample contains the nearby systems 
that the interstellar absorption and the reddening could be ignored. Moreover, the CAB 
are popular that they are usually well studied systems that we are confident to apply 
the rules of Sparke \& Gallagher (2000) for estimating the missing standard errors of 
15 (6.3\% in the list) spectroscopic parallaxes. IM Vir and HZ Com are within 60 pc. Thus, 
with a 287 pc distance, only the error of HP Aur could be doubted. Nevertheless, 
It will not effect the statistics of the whole sample. The average propagated errors 
at U,V,W for these 15 systems were computed as $\delta U=\pm 5.49$, $\delta V=\pm 4.55$ 
and $\delta W=\pm 3.81$ km/s, which are smaller than the propagated errors of nine systems 
with $\sigma_{\pi}/\pi>0.5$, but bigger than the average propagated errors of the whole 
sample : $\delta U=\pm 3.43$, $\delta V=\pm 2.92$ and $\delta W=\pm 2.42$ km/s.
    
Finally, after filling in 
the missing information in Table 1, the average standard errors on 
the proper motion components are 0.62 mas/yr in 
$\mu_{\alpha} cos\delta$ and 0.43 mas/yr in $\mu_{\delta}$ 
and the average relative uncertainty of the parallaxes 
($\sigma_{\pi}/\pi$) is $14.7\%$.  

\subsection {Radial velocities}

Unlike the proper motions and the parallaxes, which were mostly 
taken from the Hipparcos and the Tycho Catalogs, the radial 
velocities were collected one by one from the literature. 
Moreover, unlike single stars with a single radial velocity, 
the binaries and the multiple systems require the radial 
velocity for the mass center of the system ($\gamma$). 
That is, numerous radial velocity measurements are needed 
just for computing the orbital parameters together with 
the velocity of the mass center of a system. Fortunately, the 
CAB are popular so that the reliable orbital parameters had 
already been determined for many systems. However, there are 21 
systems in our list (Table 1) which are known to be binaries but 
do not yet possess determined orbital elements. For such systems, 
the mathematical mean of the measured radial velocities was adopted
as the center of mass velocity and then the standard deviation
from this mean was taken to be the error estimate. On the 
other hand, there are many systems with multiple orbit
determinations. Nevertheless, most of the multiple orbit 
determinations are not independent. That is, the data used in 
the previous determination were also used or considered in the 
later study which gives the most improved orbital elements. 
In such cases, it was preferred to use the value of ($\gamma$) 
and its associated error from the most recently determined 
orbit unless the most recent study gives unexpectedly large 
associated errors. Rarely, there are systems with 
independently determined orbital parameters. For those, the weighted
 mean of the systemic velocities ($\gamma$) and the weighted mean of the 
associated errors were used. Those systems are listed 
with the multiple reference numbers separated by commas 
after the second semicolon in the last column of Table 1. 

Different authors prefer to give different kinds of 
uncertainties associated with the published parameters 
of the orbit. In order to maintain consistency, the different 
types of uncertainties have been transformed into standard
errors since most of our data are already expressed with 
the standard errors. Except for the probable error, the other 
uncertainties (mean error, standard error, rms error and $\sigma$)
indicate the same confidence level. Therefore, they are 
transferred directly. However, when transforming the 
probable errors (PE) to the standard errors (SE), the relation 
of $PE = 0.675 SE$ was used.        

\section {Galactic space velocity components}
Galactic space velocity components ($U$, $V$, $W$) were
computed together with their errors by applying the algorithm
and the transformation matrices of Johnson \& Soderblom (1987)
to the basic data; celestial coordinates ($\alpha$, $\delta$), 
proper motion components ($\mu_{\alpha}$, $\mu_{\delta}$), 
radial velocity ($\gamma$) and the parallax ($\pi$) of 
each star in Table 1, where the epoch of J2000 coordinates 
were adopted as described in the International Celestial 
Reference System (ICRS) of the Hipparcos and the Tycho 
Catalogues. The transformation matrices use the notation 
of the right handed system. Therefore, $U$, $V$, $W$ are 
the components of a velocity vector of a star with respect 
to the Sun, where $U$ is directed toward the Galactic center 
($l=0^{o}, b=0^{o}$); V is in the direction of the galactic
rotation ($l=90^{o}, b=0^{o}$); and $W$ is towards the north 
Galactic pole ($b=90^{o}$). 
The computed uncertainties are quite small and the averages 
are $\delta U=\pm 3.43$, $\delta V=\pm 2.92$ and $\delta W=\pm 2.42$ km/s. 
By inspecting the space velocity vectors ($s=\sqrt{U^{2}+V^{2}+W^{2}}$),
only 18 (7.6\%) systems with the uncertainty of the space velocity 
bigger than $\pm15$ km/s were found. If those systems were removed 
from the sample, the average uncertainties of the components would 
reduce to $\delta U=\pm 2.4$, $\delta V=\pm2.0$, and 
$\delta W=\pm 1.8$ km/s. Thus, most of our sample stars have 
uncertainties very much smaller than the dispersions calculated.   

\subsection{The space distribution}

Before discussing the velocity dispersions and kinematical 
implications, it was decided to inspect the space distribution
of the sample CAB. Therefore, the Sun centered rectangular 
galactic coordinates $(X, Y, Z)$ corresponding to space velocity 
components $(U, V, W)$ were calculated. 
The computed coordinates are given in Table 2. The projected
positions on the galactic plane ($X, Y$ plane) and on the plane
perpendicular to it ($X, Z$ plane) are displayed in Figure 1.

\begin{figure}
\resizebox{8cm}{15.5cm}{\includegraphics*{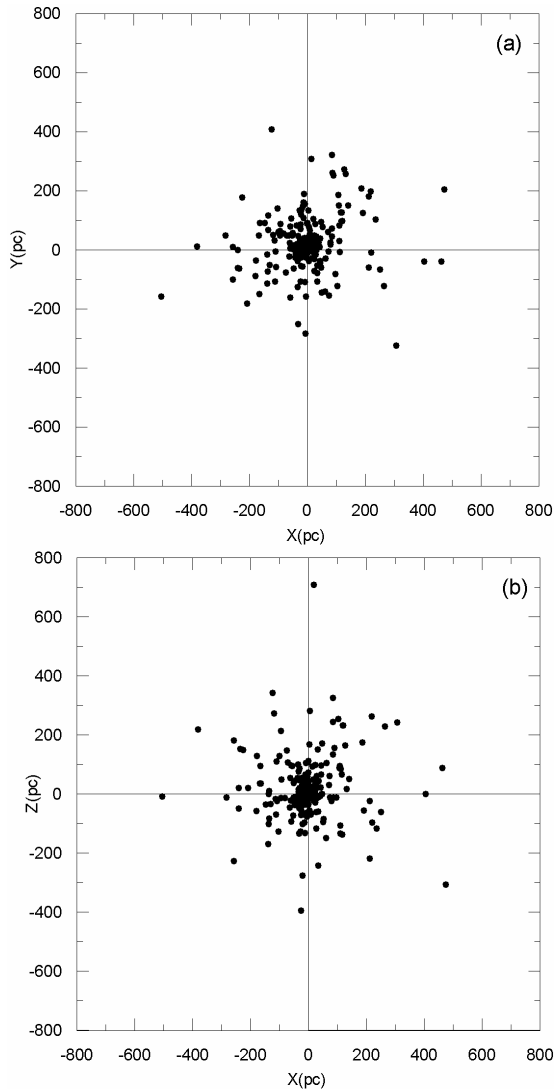}} 
%\vskip 15.5cm
%\special{bmp: c:ma968fig01.bmp x=8cm y=15.5cm}
\caption{Space distribution of sample stars on galactic plane, 
and one other of two perpendicular planes. X, Y, and Z 
directed towards galactic center, galactic rotation and 
north galactic pole.}
\end{figure}

Fig. 1 indicates that, with a median distance of 98 pc, the 
current CAB sample contains relatively
nearby systems, which can be considered as being contained
within the galactic thin disk. They can also be accepted as 
almost homogeneously distributed in all directions as they are
seen from the Sun.

\subsection{Galactic differential rotation correction}

The high accuracy of the $U$, $V$, $W$ velocities motivated us to 
investigate the effect of the differential galactic rotation
to the $U$, $V$, $W$ velocities. The effect of the galactic differential 
rotation is proportional to the distance of stars from the Sun in 
the galactic plane, that is, the W velocities are not affected in 
the first approximation which assumes stars are on the galactic plane.
Since all of the systems are relatively nearby, the first order 
correction described in Mihalas \& Binney (1981) was announced 
to be smaller than the uncertainties of $U$ and $V$ by Eker (1992) 
for the 146 CAB which  also exist in the present list. 
Nevertheless, there was no harm in applying the correction 
even if it is negligible, as Eker (1992) explained.

Since the largest uncertainty of the input data appears to 
be with the parallax measurements, the uncertainty of the 
distance contributes the most to the uncertainties of the 
$U$, $V$, $W$ velocities when compared to the contributions of 
proper motions and radial velocities. With the greatly 
improved astrometric data of Hipparcos which produces 
reliable parallax measurements up to 500 pc, the uncertainties 
in the $(U, V, W)$ space motions are greatly reduced 
(nearly five times) compared to the data used by Eker (1992). 

Using the space distribution in $X, Y$ plane in Fig. 1, 
the first order galactic differential correction contributions
to the $U$ and $V$ space motions were computed as described in 
Mihalas \& Binney (1981). Then, star by star, they were 
compared to the uncertainties of the $U$ and $V$ computed. 
It was not unexpected to see 128 stars ($54\%$) in our list 
with the effect of galactic differential rotation being bigger 
than the uncertainty of $U$ component of the space velocity. The
effect on the $V$ component is rather small, therefore, there are 
only three CAB with the effect being bigger than the uncertainty 
of $V$.  Nevertheless, it seems evident that the first order galactic
differential rotation correction is necessary for most of the stars 
in our sample. Therefore, the first order correction of galactic 
differential rotation was applied to all of the stars in the present
sample. The corrected $U$, $V$, $W$ are given in Table 2, together 
with the propagated standard errors.

\subsection{Thick disc and halo binaries}

The number of metal poor binaries in our sample was also determined by 
using the kinematical parameter ${\it f\/}=(1/300)(u^{2} + 2.5 v^{2}
+3.5 w^{2})^{1/2}$ suggested by Grenon (1987) and Bartkevicius et al.\ 
(1999). Here, the $u, v, w$ velocities represent a space velocity with 
respect to the LSR. The $(u, v, w)$ velocities are obtained by adding the 
velocity of the Sun with respect to the LSR to the $(U, V, W)$ velocities 
of stars with respect to the Sun. 
The values of $(U, V, W)_{\odot}=(9, 12, 7)$ km/s (Mihalas \& Binney 1981)
were used in this transformation. Statistically, the stars with 
${\it f\/}\leq0.35$ belong to the thin disc, the stars with 
$0.35 < {\it f\/}\leq1.00$ belong to the thick disc. The stars with $f>1$ 
belong to the halo. Consequently, the vast majority ($92\%$) of our sample 
are thin disc stars. The thick disk stars are less composing about $7\%$ 
of CAB in our sample. Only one binary star, HD149414 is a halo star 
according to its space motions (kinematically). The spectroscopic metal 
abundance ($[m/H]=-1.40$ dex) of this star given by Latham et al.\ (1988) 
confirms the classification based on the kinematical criteria.
The Hipparcos parallax of this star gives the distance of 48 pc, 
so it appears to be a halo binary in the solar neighborhood.
This binary has a long period (133 days) and a eccentric orbit
(Mayor \& Turon 1982). It is interesting that 
Buser, Rong, \& Karaali (1999), and Siegel et al.\ (2002) 
found that the $6\%$ of the solar neighborhood stars belong 
to the thick disc population, which is an almost identical 
ratio to our CAB sample.

\section{discussion}
\subsection{General outlook}

The distribution of the corrected $U$, $V$, $W$ velocities on the 
$(U, V)$ and $(W, V)$ planes are displayed in Fig. 2. At first glance, 
the general look of the current $(U, V)$ diagram (Fig. 2a) appears to
have similar characteristics to the $(U, V)$ diagram of Eker (1992)
of the same sample but fewer stars (146) with much lower accuracy. 
Quantitatively speaking, the average motion of the current sample with
respect to the Sun is $(U, V, W)=(-13.5, -19.7, -8.1)$ km/s having the
dispersion of (37.3, 26.0, 19.4) km/s with respect to the LSR,
which are indeed close to the values of Eker (1992): 
$(U, V, W)=(-10, -20, -7)$ km/s and the dispersions of
(37, 27, 23) km/s. Later, Aslan et al. (1999) also studied 
the kinematics of the 178 CAB using the Hipparcos astrometric
data. The shape and the distribution characteristics of
the $(U, V)$ diagram of Aslan et al. (1999) also has a similar
appearance to the mean motion of $(U, V, W)=(-11.8, -20.5, -6.4)$ 
km/s relative to the Sun and $(35.8, 22.4, 18.2)$ km/s dispersions 
in the space velocities with respect to LSR.   

The similar appearance, the similar mean velocity and the
similar distribution of the current sample on the $(U, V)$ does 
not seem to display any advantage of increased accuracy and 
increased number over the previous studies. However, as soon as
our first $(U, V)$ diagram was produced, the $\gamma$ shaped
concentration of $(U, V)$ velocities near the LSR (See Fig. 2a) was
noticed. Such a concentration of kinematically young systems is
not noted by neither Eker (1992) nor Aslan et al. (1999). However,
the $\gamma$ shaped concentration is very clear in Fig. 2a.
The concentration of the young systems is also noticeable 
on the $(W, V)$ plane (Fig. 2b) but in a rather spherical shape.

\begin{figure}
\resizebox{8cm}{15.5cm}{\includegraphics*{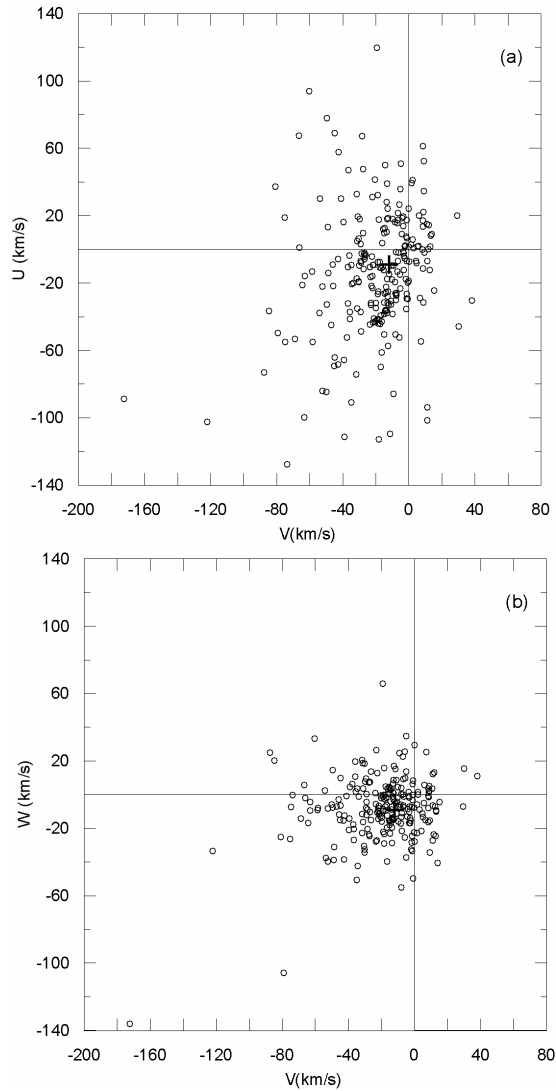}} 
%\vskip 15.5cm
%\special{bmp: c:/pctexv4/ma968fig02.bmp x=8cm y=15.5cm}
\caption{Velocity dispersions of CAB sample (a) on U, V plane, 
(b) on W, V plane. The velocities are heliocentric. 
The position of LSR is marked by +.}
\end{figure}

Containing stars from F to M spectral types on the main 
sequence, together with the evolved G and K giants and
even with the super giants, the studied samples of CAB 
happens to be very heterogeneous. On the other hand, the
orbital periods of the binaries in the sample range from 
fractions of a day to more than 300 days. This may mean that 
there are different evolutionary paths (Plavec \ 1968; Thomas \ 1977)
indicating different ages existing together among the sample stars.
As Eker (1992) investigated and Aslan et al. (1999) announced, 
there could be no significant kinematical differences between 
the sub-samples of the CAB except some indication of the main 
sequence RS CVn systems tending to have smaller velocity
dispersions implying smaller ages.    

The difficulty of separating kinematically young and old
populations in the velocity space alone is obvious. The
dispersions increase with age but there are always some 
stars left near the LSR. It is therefore not safe to pick 
stars randomly near the LSR and then to form a kinematically
young group with them. The classical approach would be to 
form the sub groups according to certain objective criteria 
at first, then to investigate and to compare the dispersions 
among the sub groups.
	
However, the concentration of velocities around $(U, V)=(17, -8)$, 
$(U, V)=(-4, -26)$, $(U, V)=(-37, -14)$, and $(U, V)=(0, 0 )$ km/s 
perhaps reflect some kinds of group motions of the stars in the 
solar vicinity. Eggen (1958a-b, 1989, 1995) and Montes et al. (2001a) 
discuss the possible moving groups (Local Association, Ursa Major,
Castor, IC 2391, and Hyades), which might cause the concentration 
of space velocities as described above. Therefore, as a first 
step before examining the classical sub groups, it was decided 
to determine the MG members of our sample and then investigate 
if the $\gamma$ shaped concentration is caused by them. Moreover,
the membership of one of the known MG would be an objective 
criterion to discriminate the kinematically young population of 
the present CAB sample.  

\subsection{Members of MG among CAB}

The kinematical criteria originally defined by Eggen (1958a, b, 
1989, 1995) for determining the possible members of the best documented 
moving groups are summarized by Montes et al. (2001a,b). Basically, 
there are two criteria: 

(i) The proper motion criterion, which uses the ratio ($\tau$/$\nu$) as a 
measure of how the star turns away from the converging point, where the 
$\nu$ and the $\tau$ are the orthogonal components of the proper motion 
($\mu$) of a test star. The component $\nu$ is directed towards the 
converging point and the $\tau$ is perpendicular to it on the plane of 
the sky. A test star becomes a possible member if 
$(\tau/\nu)<(0.1/sin \lambda)$, where the $\lambda$ is the angle 
corresponding to the arc between the test star and the converging point. 

(ii) The radial velocity criterion, which compares the observed 
radial velocity ($\gamma$, the center of mass velocity) of the 
test star to the predicted mean radial velocity
$V_{p}= V_{T}cos\lambda$, where $V_{T}$ is the magnitude of the 
space velocity vector representing the MG as a whole.
The test star is a possible member if the difference between 
$\gamma$ and $V_{p}$ is less than the dispersions of the radial 
velocities among the stars in the MG. 

Fulfilling one of the criteria makes the test star a possible 
member. Fulfilling both criteria, however, does not guarantee the 
membership. This is because there is always a possibility that the 
same velocity space is occupied by the MG members and the non 
members. Further independent criteria implying a common origin and 
same age as the member stars may be investigated in order to 
confirm the true membership.     

The parameters of the five best documented MG and
the possible membership criteria of each of them have 
been summarized in Table 3. The criteria have been applied 
one by one to all stars in our CAB sample and 95 systems 
out of 237 were found to be satisfying at least one of the 
criteria for one of the MG in Table 3. Those potential 
candidates are marked on Table 2 indicating the number 
of criteria fulfilled (1 means only one criterion, 
2 means both criteria were satisfied) and the name of 
the MG involved. Some already known members are also 
marked on a separate column for a consistency check 

\setcounter{table}{2}
\begin{table*}
\caption{Parameters of best documented moving groups and 
possible membership criteria.} 
\begin{tabular}{lcccccc}
\hline
      Name &        Age &  (U, V, W) &    $V_{T}$ &  C.P. & $Sin \lambda (\tau/\nu)$ & $\gamma-V_{p}$ (km/s) \\
           &      (Myr) &     (km/s) &     (km/s) &   ($\alpha^{h}$, $\delta^{o}$) &            &            \\
\hline
Local Association &   20 -- 150 & (-11.6,-21.0,-11.4) &       26.5 & (5.98,-35.15) &       $<$0.2 &       $<$5.5 \\
(Pleiades, a Per, M34, &            &            &            &            &            &            \\
$\delta$ Lyr, NGC 2516, IC2602) &            &            &            &            &            &            \\
IC 2391 Supercluster &    35 -- 55 & (-20.6,-15.7,-9.1) &       27.4 & (5.82,-12.44) &       $<$0.1 &         $<$7 \\
 (IC 2391) &            &            &            &            &            &            \\
 Castor MG &        200 & (-10.7,-8.0,-9.7) &       16.5 & (4.75,-18.44) &       $<$0.1 &         $<$8 \\
Ursa Major Group &        300 & (14.9,1.0,-10.7) &       18.4 & (20.55,-38.10) &       $<$0.1 &        $<$8 \\
(Sirius Supercluster) &            &            &            &            &            &            \\
Hyades Supercluster &        600 & (-39.7,-17.7,-2.4) &       43.5 & (6.40,6.50) &       $<$0.1 &        $<$10 \\
(Hyades, Praesepe) &            &            &            &            &            &            \\
\hline
\end{tabular}  
\end{table*} 

After all of the possible MG members were determined, the sample
was divided into two groups. The one which contains the possible 
MG members is called `MG' and, the other, which contains the rest 
of the sample is named `field stars'. The $(U, V)$ diagram of these 
groups are compared in Fig. 3. The $\gamma$ shaped concentration 
which was noticed on the $(U, V)$ diagram of the whole sample (Fig. 2a) 
shows itself more clearly in Fig. 3a after the removal of stars 
which fail to be a possible member of any of the five MG in Table 3. 
The smooth distribution (Fig. 3b) with a larger dispersion 
of the field stars is also clear on the comparison with the
whole sample (Fig. 2a) and the possible MG members (Fig. 3a).
Comparison of these two groups on the $(W, V)$ diagram are 
displayed in Fig. 4.     

\begin{figure}
\resizebox{8cm}{15.5cm}{\includegraphics*{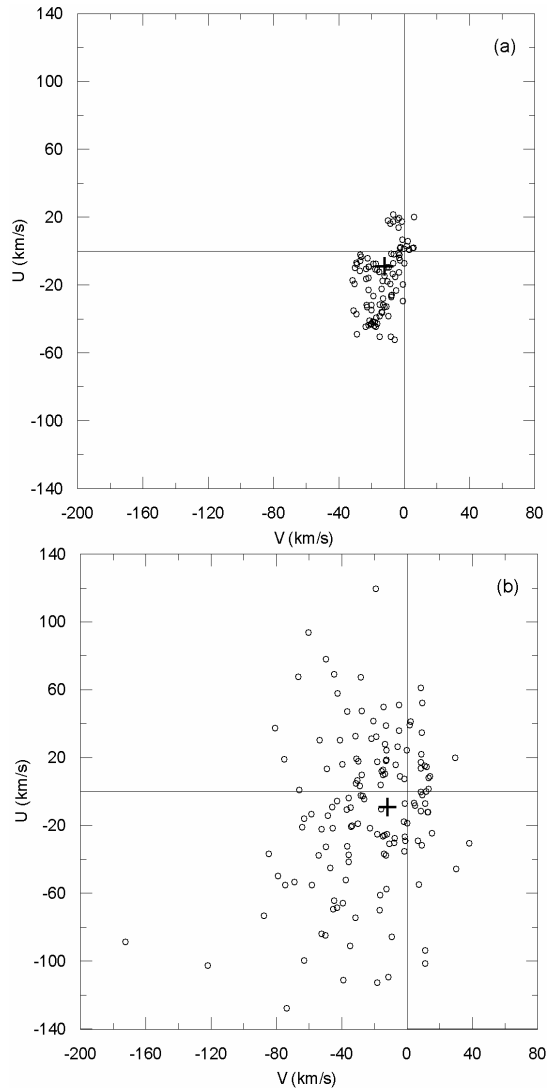}} 
%\vskip 15.5cm
%\special{bmp: c:/pctexv4/ma968fig03.bmp x=8cm y=15.5cm}
\caption{Distribution of (a) possible MG members and (b) 
field stars on the $(U, V)$ diagram.}
\end{figure}

\begin{figure}
\resizebox{8cm}{15.5cm}{\includegraphics*{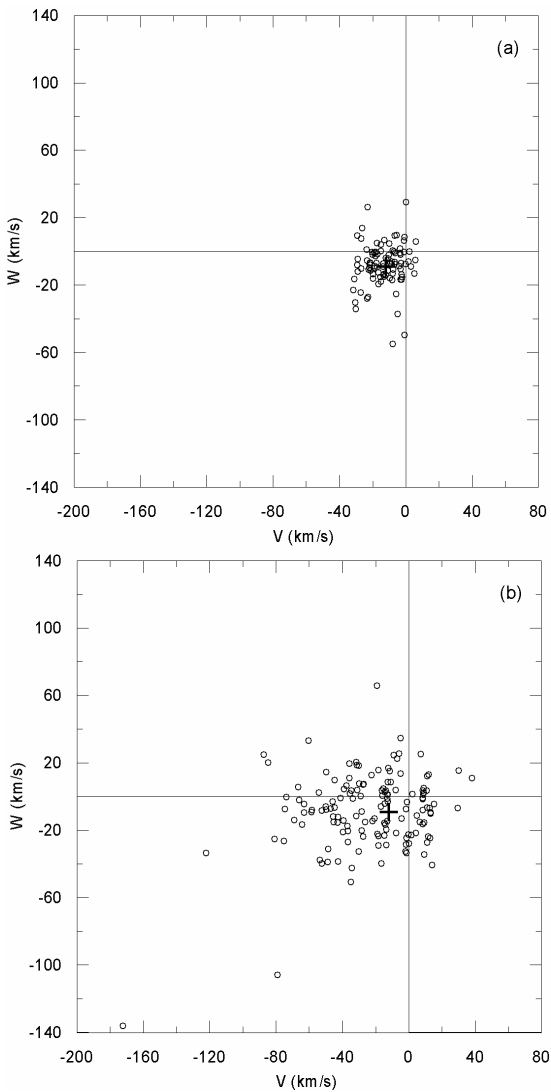}} 
%\vskip 15.5cm
%\special{bmp: c:/pctexv4/ma968fig04.bmp x=8cm y=15.5cm}
\caption{Distribution of (a) possible MG members and (b) 
field stars on the (W, V) diagram.}
\end{figure}

The kinematical differences between the two groups of CAB 
can be shown numerically if their mean motions and dispersions
are compared. The `MG' has a mean motion of $(U, V, W)=
(-16.9, -13.5, -7.6)$ km/s with the dispersions of (20.6, 9.8, 12.8) 
km/s while the `field stars' appear with a mean motion of
$(U, V, W)=(-11.2, -24.0, -8.4)$ km/s and the dispersions 
of (45.4, 32.9, 22.9) km/s. According to Wielen (1977),
$\sigma_{U}$ = 20.81,  $\sigma_{V}$= 9.76, $\sigma_{W}$= 12.74 km/s
velocity dispersions indicate a kinematical age of 950 Myr,
which is slightly bigger than the known ages of the MG given 
in Table 3. This is because the dispersion of stars was computed
with respect to the LSR. However, true age would be less if the
true dispersion point of each group is considered. Considering the 
fact that some of the possible moving group members are not really 
members, this age can be treated an upper limit. On the other hand,
the kinematical criteria to form the `MG' group chooses only a 
limited number of young binaries, there can be binaries left in the 
`field stars' younger than 950 Myr. Thus, this age (950 Myr) cannot 
be considered as a lower limit for the `field stars' which are 
found to have 3.86~Gyr age from the dispersions. There can be stars 
much younger and older than this average age among the `field stars'.       

On careful inspection of Fig. 3b and Fig. 4b, one may notice the distinct
holes left in the centers of the distributions after the possible MG 
members removed. This confirms the fact implied by the term `possible', 
and suggests a substantial amount of the MG stars are really not MG members. 
Any individual systems being older than the common age of the MG could be 
selected out as non-members with the ages predicted by the stellar evolution, 
but this process too does not guarantee to remove all of the 
non-members since there is still a possibility that a field star, with a 
similar age as the MG, occupies the same velocity space fulfilling the 
kinematical criteria to be a possible member. Nevertheless, our prime concern 
is to divide current CAB sample into two distinct age groups in order to 
compare the physical parameters then investigate the reasons behind if there 
is any noticeable difference. Although, both the `MG' and `field stars' are 
not very homogeneous to represent two different ages, we found current grouping 
satisfactory for this study.

\subsection{Comparing `field stars' and `MG'}

The physical parameters of the chromospherically active 
binaries are listed in Table 4. The columns are self 
explanatory and indicate the spectral type, SB 
(indicating single or double lined binary or whether 
within a multiple system), orbital period, eccentricity, 
mass ratio, mass of the primary, mass function, and radii
of components. The data were collected primarily from 
the same literature where the radial velocities were taken. 

Intending to compile binaries according to known evolutionary 
stages of luminosity classification, the whole sample has been 
divided into three groups. The first group is called `G' which 
contains binaries with at least one component being a giant. 
A giant classification in the spectral type, if it exists, or 
otherwise, one of the radii being six solar radii or bigger, 
were accepted as criteria to form the `G' group. The group of 
the sub giants `SG' were formed from the rest of the sample 
with a similar criteria; a sub giant classification in the 
spectral type, or at least one component being bigger than 
two solar radii. After forming the giants and sub giants, 
the rest of the sample is called main sequence symbolized 
with `MS'. All three groups contain almost equal numbers of 
`MG' and `field stars'.

In the first step, the mass and period distributions among 
those three groups were studied. The result confirms common 
knowledge that the massive systems are likely to be found in 
the group of the `G' and the less massive systems are likely 
to be found among the `MS' group, so it is not displayed. 
However, it is of interest to display the period distribution 
(Fig. 5) among the G, SG, and MS systems. The `SG' group shows 
nearly a normal distribution with the peak at six days and a range 
of orbital periods from 0.79 to 50 days. The group of `G' 
prefers not only more massive systems but also the systems with 
the longest orbital periods. According to Fig. 5, systems containing 
a giant star prefer an orbital period of 10 days or longer, but 
rarely shorter periods. Notice the sharp decrease of the short period 
binaries in the `G' group. The `MS' systems are mostly less 
than 10 days down to the shortest period of 0.476533 days. Our sample
does not have many shorter periods because CAB are detached systems. 
Much shorter periods are common among the contact (W UMa) and semi
contact ($\beta$ Lyrea) binaries. It is interesting to note that 
the range of `MS' periods covers quite a range of the most preferred `G' 
group periods with a smooth decrease. This decrease may well be due to the 
selection effect that main sequence long period systems are harder to be 
noticed than the long period binaries with a giant or two. However, similar 
selection effect cannot be true for the decrease of the `G' 
systems towards the shorter periods. 

\begin{figure}
\resizebox{7cm}{11cm}{\includegraphics*{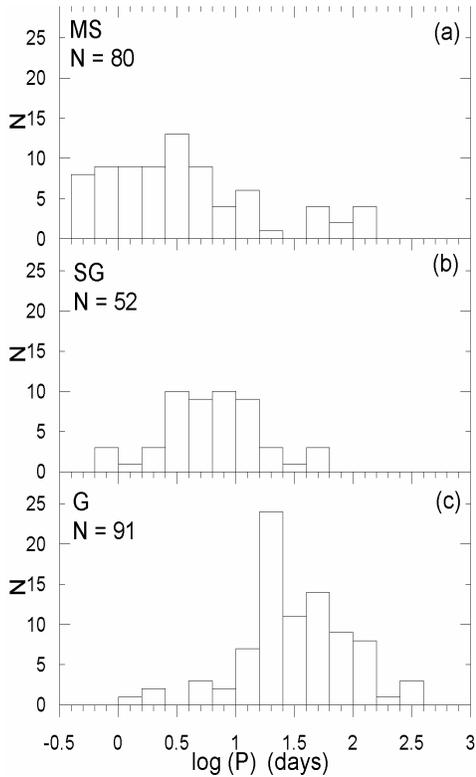}} 
%\vskip 11cm
%\hskip .5cm
%\special{bmp: c:/pctexv4/ma968fig05.bmp x=7cm y=11cm}
\caption{Histogram of period distribution among binary systems 
containing MS (main sequence), SG (sub giant), and giant (G). 
All groups contain almost equal numbers of younger `MG' and 
older field binaries.}
\end{figure}

\begin{figure}
\resizebox{7cm}{8cm}{\includegraphics*{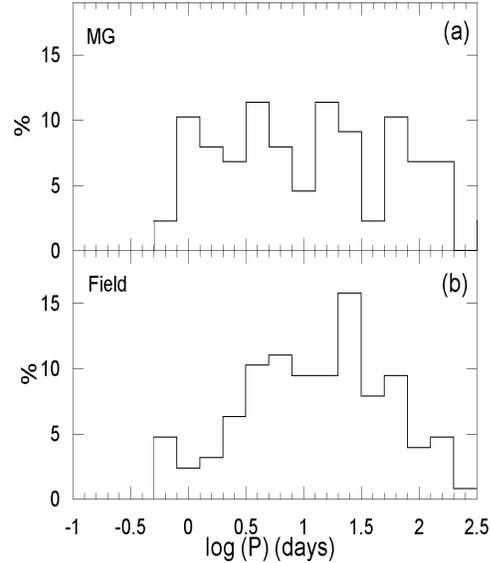}} 
%\vskip 8cm
%\hskip .5cm
%\special{bmp: c:/pctexv4/ma968fig06.bmp x=7cm y=8cm}
\caption{Comparison of period histograms of (a) MG and 
(b) field stars.}
\end{figure}

\begin{figure*}
\resizebox{15cm}{11cm}{\includegraphics*{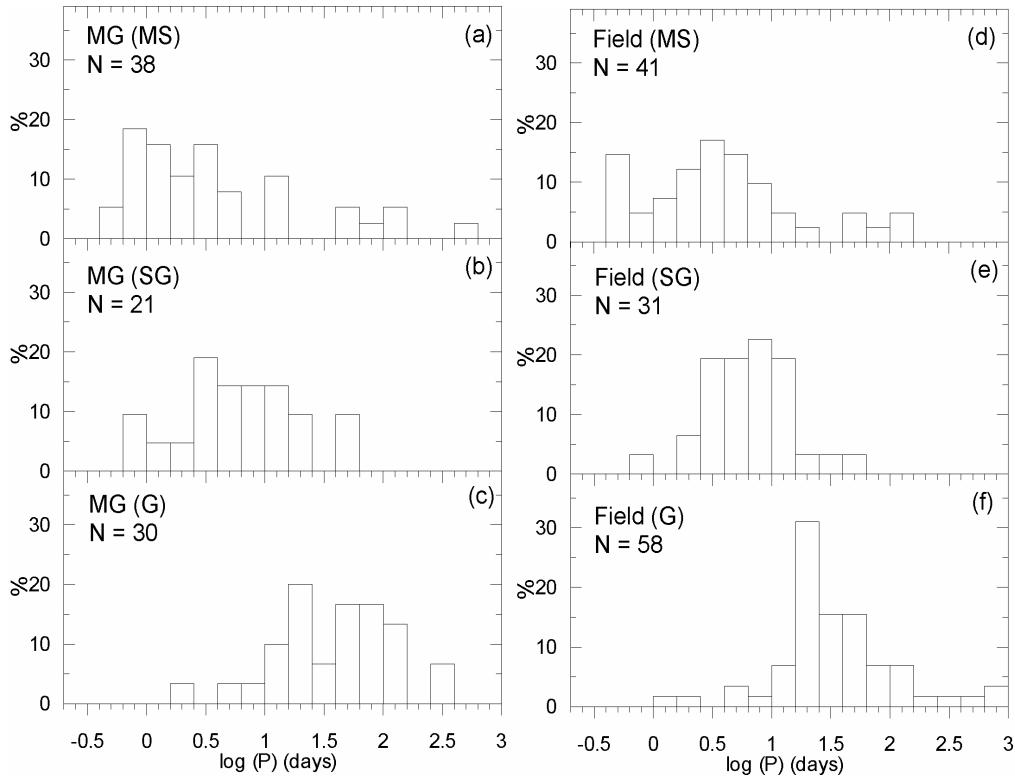}} 
%\vskip 11cm
%\special{bmp: c:/pctexv4/ma968fig07.bmp x=15cm y=11cm}
\caption{Comparison of period histograms of sub groups 
between MG and field binaries. (G) at least one component 
is giant, (SG) sub giant, and (MS) both components on the 
main sequence.}
\end{figure*}

Fig. 6 compares the orbital period distributions between 
the kinematically younger (MG) and older (field) populations
in our sample. Both groups have about the same range of 
orbital periods. However, the younger MG group shows a 
rather smoother distribution, without a distinct peak,
contrasting with the older population, which shows a
peak of a gaussian at 11.3 ($logP=1.053$) days.
At first, the composition rates of G, SG, and MS systems in both 
groups were investigated. There are 88 systems in the younger 
population in Fig. 6a which is composed of $34\%$ G, $24\%$ SG, and 
$42\%$ MS systems. On the other hand, there are 127 of the stars in 
the older population in Fig. 6b which is composed of $43\%$ of G, 
$25\%$ of SG, and $32\%$ of MS systems. There is not much 
difference in the distributions of the subgroups between the 
two groups. Therefore, the period preferences of the sub groups 
(G, SG, and MS) alone cannot not explain the displayed difference 
between Fig. 6a and Fig. 6b. Nevertheless, the decrease in the 
number of systems in the longer and shorter periods in field stars 
(the older sample) may be an effect in the binary evolution.

According to Demircan (1999), mass loss from a binary is associated 
with the momentum loss causing the decrease of the semi-major axis of 
the orbit. A shrinking orbit forces the orbital period to decrease. 
Fig. 6 appears to support this scenario. This is because, assuming that 
the `field stars' have a similar period distribution as `MG' at the origin 
when they were younger, the number decrease of longer period systems could 
be interpreted with the above prediction. However, the number decrease of 
short period systems appears to contradict the scenario. That is, normally 
one expects to count more systems with shorter periods among the older 
binaries if orbital periods decrease during evolution. However, it should 
not be forgotten that the binaries in our sample are all detached systems. 
Apparently, the period decrease and radius increase in the evolution 
changed those short period systems into contact or semi contact form, 
thus they are no longer in our sample and we see their number decreased 
relative to the original population. Therefore, the number decrease of 
the short period systems in Fig. 6b also supports the prediction of period 
decrease in the binary evolution. 

By comparing the period histograms of the G, SG, and MS systems between the
MG and the field stars, Fig. 7 also presents evidence of decreasing
orbital periods during the binary evolution. It is noticeable that 
the histogram of G systems for the field stars shows a sharp peak 
at 20 ($logP=1.3$) days. There is a sharper decrease towards the 
shorter periods. Such a sharp decrease is not visible in the young
population (G systems of MG). This sharp decrease could be caused 
by the missing systems which are no longer on the list; due to evolution
they became contact or semi contact systems. The shifting of the
peak of the normal distribution towards the shorter periods as an evidence 
of the orbital period decrease is clearly visible in the comparison of 
the G groups; perhaps among the SG. Nevertheless, the opposite, that is, 
the peak of the distribution of f,eld (MS) systems appears to be at longer 
periods with respect to the peak of the MG (MS) systems. However, considering 
the fact that evolving into contact, or semi-contact configuration is most 
likely among the short period MS systems rather than G systems, therefore it 
could be normal to see the peak moving towards the longer periods 
in the statistics of the MS systems. The MS systems causing a peak at 
around the one day period in the MG group must have evolved to contact 
or semi contact configurations so that the number of such systems appears 
to be less in the field stars. Therefore, the peak appears to have moved 
towards the longer periods for field (MS) binaries. 

One may ask why the peak of field (MS) binaries indicates 
a shorter period than the peak of field (G) binaries if evolution
to contact configuration is effective for up to 10 days, which 
is indicated by the histogram of the field (G) binaries. Here, 
we must remember that neither the younger (MG) nor the older (field) 
group are very homogeneous. There could be older binaries among the 
possible MG members, so they are called possible, and there could be
many young binaries among the field stars. The kinematical criteria
only select possible MG members. It is possible that 
unselected stars could be young systems but not satisfy the MG criteria. 
This complication, however, is not to such a degree that despite this 
heterogeneous nature, the period shortening effect of the 
binary evolution is perceptible on our histograms. It is a challenge 
for future studies to select the older systems from the possible MG 
members and select the younger systems from the field stars for a 
better comparison of the younger and the older groups of binaries. 

\setcounter{table}{4}
\begin{table}
\caption{Kinematical ages of period sub groups in field stars.}
\center
\begin{tabular}{cccc}
\hline
$log (P)$ (days)&    N & $\sigma_{tot}$ (km/s) &  Age (Gyr) \\
\hline
(0.0 -- 0.8]  &   48 &      61.38 &       6.69 \\
(0.8 -- 1.7]  &   59 &      53.15 &       5.19 \\
(1.7 -- 3.0]  &   23 &      40.99 &       3.02 \\
\hline
\end{tabular}  
\end{table}  

Orbital periods decreasing with age are confirmed by the 
kinematical data. The older population (field stars) has been
divided into three period ranges (Table 5) and the space 
velocity dispersions and kinematical ages were calculated for
the short ($logP\leq0.8$), intermediate ($0.8<logP\leq1.7$) and
the long period ($1.7<logP\leq3.0$) systems. The increase of the
dispersions, implying older ages, towards the shorter periods 
appears to support the period histograms, that is, the orbital
period decrease must be occurring during the binary evolution.

\setcounter{figure}{8}
\begin{figure*}
\resizebox{15cm}{11cm}{\includegraphics*{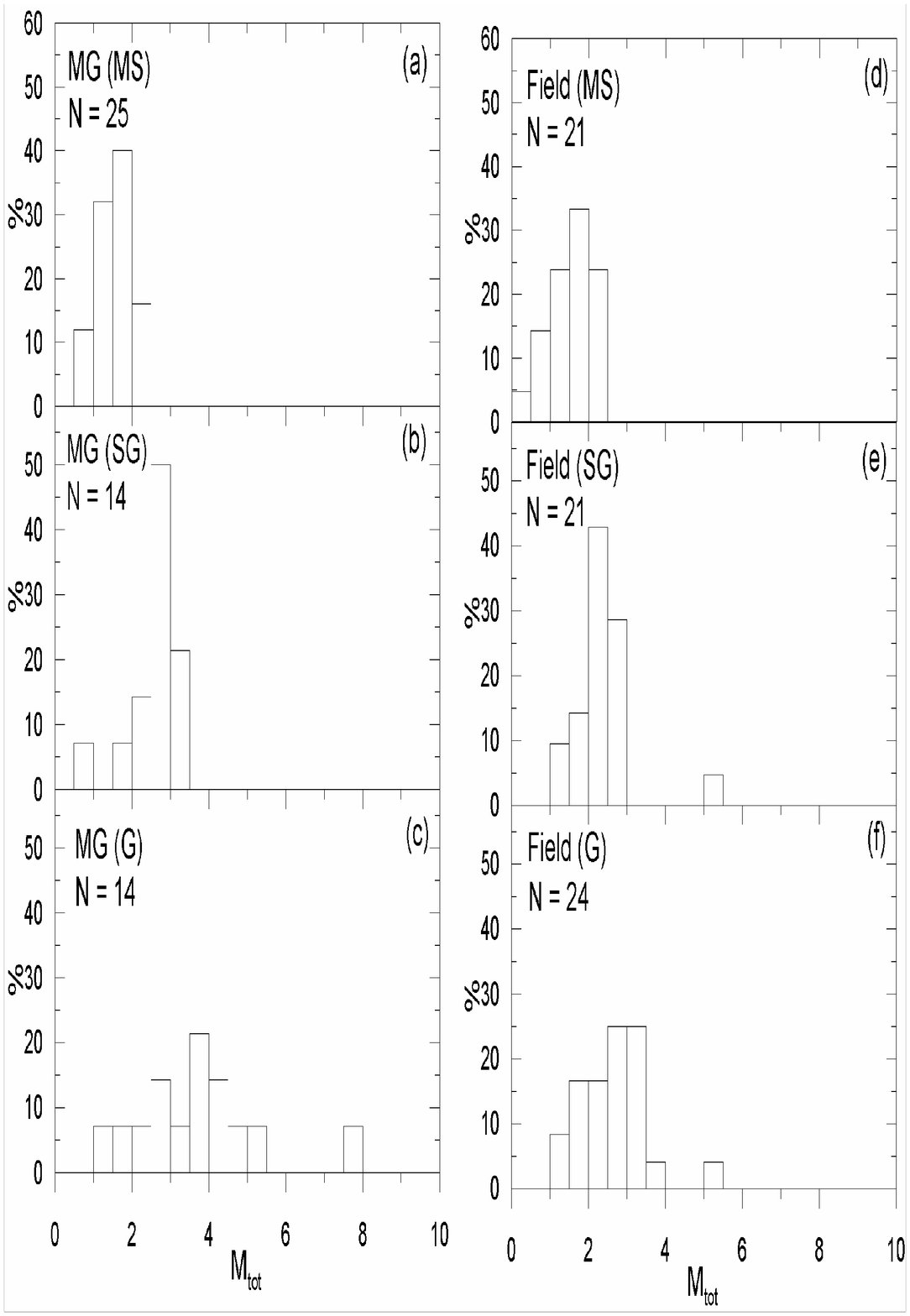}} 
%\center
%\vskip 11cm
%\special{bmp: c:/pctexv4/ma968fig09.bmp x=15cm y=11cm}
\caption{Comparison of total mass histograms of sub 
groups. (G) at least one component is giant, (SG) sub 
giant, and (MS) both components on the main sequence.}
\end{figure*}

The period decrease due to angular momentum loss requires that
the total mass of the binaries must be decreasing through 
the magnetically driven winds in CAB components (Demircan\ 1999). 
The distribution of binaries with respect to total masses 
($M_{h}+M_{c}$) in MG and field binaries are compared in Fig. 8.
The expectation was to be able to see the peak of the older
group shifting towards the smaller values with respect to the 
peak of the younger group. However, the opposite is presented 
in Fig. 8. Contrary to the peak points, the tails of the histograms
support the prediction of the total mass decrease of binaries. 
Indeed, the gradual decrease of the tail for the stars changed to 
a sharper decrease in the field stars towards the massive
systems. That is, the big mass systems in the young population 
changed to smaller mass systems in the older population. Similarly, 
sharp number decrease of the younger population (MG) towards the less 
massive systems changed to a rather gradual decrease in the older 
population (field stars). Both indicates mass decrease in the binary 
evolution. However, the heterogeneity and the evolution into contact 
or semi contact configuration complicates the histograms, making the 
interpretation of the peaks more difficult. Therefore, the young and 
old groups of Fig. 8 are separated to compare the G, SG, and MS systems 
in Fig. 9. The decrease of the total mass, and therefore the shifting 
of the peak of the distribution towards the smaller masses, became 
noticeable in the comparisons of the G and SG systems but not very 
clear in MS systems. However, it may be interesting to note that the 
low mass tail of the MS systems of field stars is longer compare to 
the tail of MS systems of MG.

\setcounter{figure}{7}
\begin{figure}
\resizebox{7cm}{8cm}{\includegraphics*{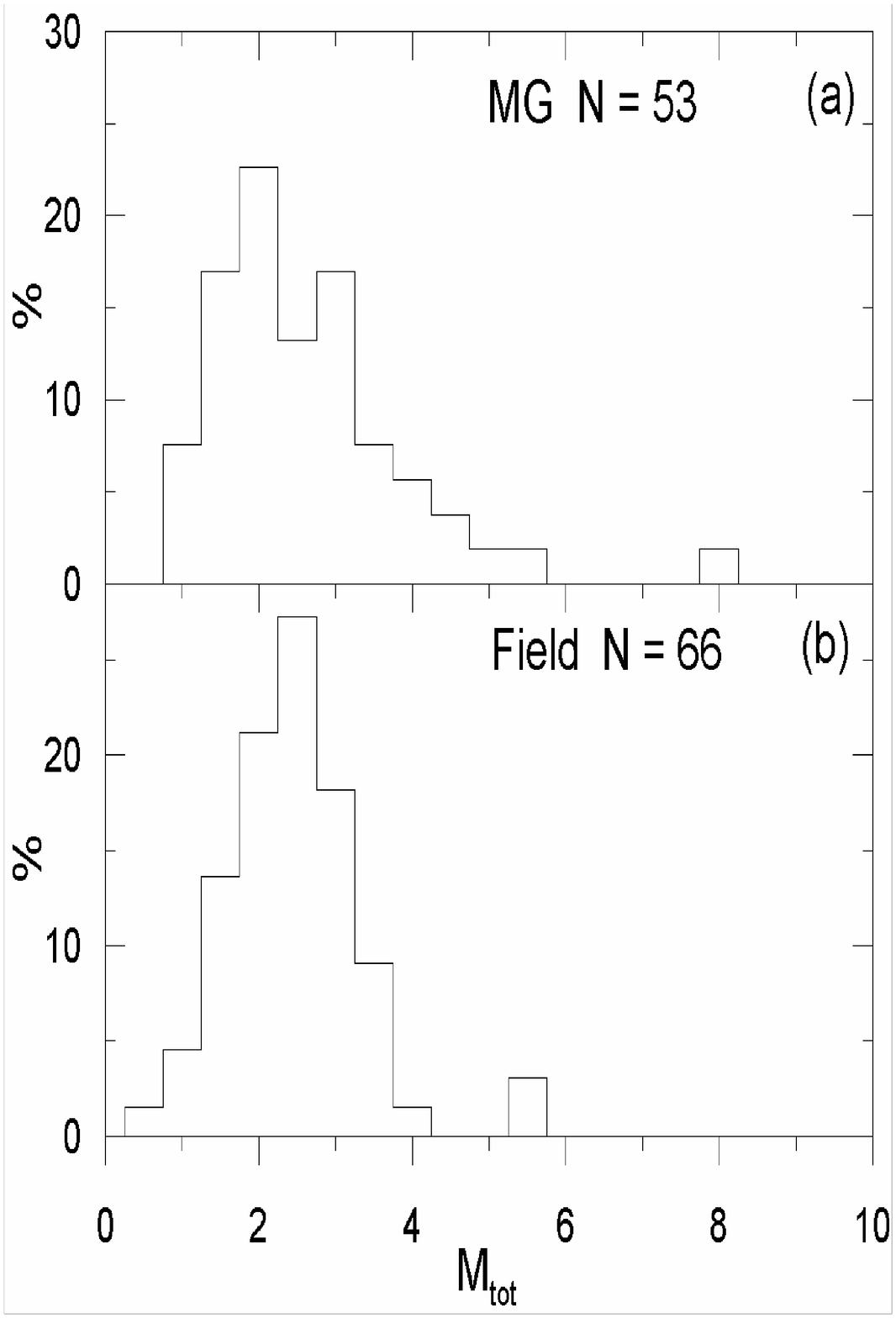}} 
%\vskip 8cm
%\hskip .5cm
%\special{bmp: c:/pctexv4/ma968fig08.bmp x=7cm y=8cm}
\caption{Comparison of the total mass ($M_{h}+M_{c}$) 
histograms of (a) MG and (b) field stars.}
\end{figure}

Fig. 10 compares the eccentricity histograms of the MG and the 
field stars. The field stars have a slightly higher peak at 
$e=0$ (circular orbits) but high eccentricity orbits exist 
at a similar level in both of the populations. The circularization
of binary orbits are expected to be faster at shorter period
orbits. Since both groups contain long period orbits, it is 
normal to see eccentric orbits in both groups. However, it is 
interesting to see a decrease in the relative number for the 
slightly eccentric orbits ($e\sim0.1$) in the field stars. 

\setcounter{figure}{9}
\begin{figure}
\resizebox{7cm}{7.9cm}{\includegraphics*{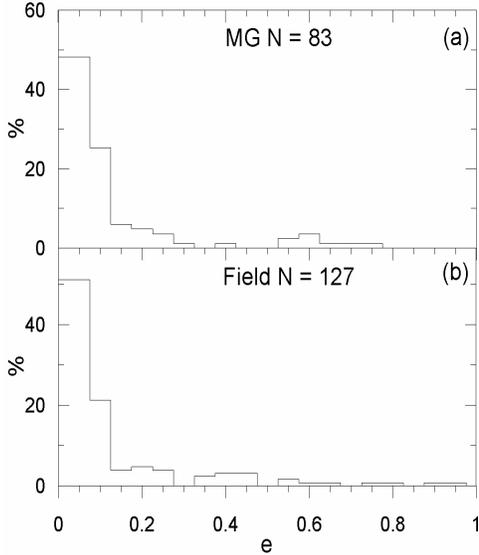}} 
%\center
%\vskip 8cm
%\special{bmp: c:/pctexv4/ma968fig10.bmp x=7cm y=7.9cm}
\caption{Comparison of eccentricity histograms between 
MG and field binaries.}
\end{figure}

\begin{figure}
\resizebox{7cm}{7.9cm}{\includegraphics*{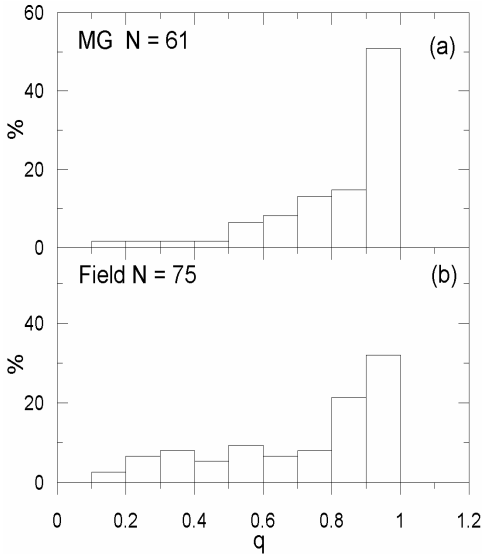}} 
%\center
%\vskip 8cm
%\special{bmp: c:/pctexv4/ma968fig11.bmp x=7cm y=7.9cm}
\caption{Comparison of mass ratio ($q = M_{2}/M_{1}$, 
where $M_{1}$ is primary and $M_{2} <M_{1}$) histograms 
between MG and field binaries.}
\end{figure}
In order to compare the mass ratio between MG and field binaries, 
the mass ratio histograms in Fig. 11 were produced. The mass ratio 
$q=M_{2}/M_{1}$, where $M_{1}$ is primary and $M_{2}< M_{1}$, is 
defined for Fig. 11. The difference is clear, in that the peak at 
$q=1$ decreased and the number of low mass ratio binaries 
increased among the field stars. This is expected because during 
the binary evolution the mass ratio of $q=1$ must decrease 
towards the smaller values. Because of the problems defining the 
mass ratio ($M_{2}/M_{1}$ or $M_{h}/M_{c}$) and the changing role 
and temperature of the components during binary evolution (a hotter 
component in the MS may become cooler as it evolves to sub giant and 
giant), the interpretation of Fig. 11 is not easy. Therefore, only 
the possible decreasing of the mass ratio through the evolution 
from MG to field binaries is pointed out here.                         

\section{Acknowledgments}
This research has been made possible by the use of the SIMBAD database, 
operated at CDS, Strasbourg, France, and the ARI database, Astronomisches 
Rechen-Institut, Heidelberg, Germany. Thanks to D. Latham for providing 
private communications. We like to thank, TUBITAK, Turkish Research 
Council and the Research Foundation of \c{C}anakkale Onsekiz Mart 
University for their partial support on this research. Finally, we 
would like to thank the anonymous referee for his/her valuable comments.

\bsp
\setcounter{table}{0}
\begin{table*}
\begin{minipage}{185mm}
\caption{Hipparcos astrometric and radial velocity data of the Chromospherically Active Binaries.}
{\scriptsize
\begin{tabular}{clccrrrrrrr}
\hline
ID&Name&HD&HIP&$\alpha$(2000)~~~&$\delta$(2000)~~~&$\mu_{\alpha}cos\delta$~~~&$\mu_{\delta}$~~~~~~&  $\pi$~~~~~ &$\gamma$~~~~~~& References \\
        &        &        &            
& (h~m~s)~~~&($^{o}$~~$'$~~$''$)~~~&(mas/yr) &(mas/yr) &(mas)~~~~&(km/s)~~~&            \\
\hline
         1 &     BC Psc &         28 &        443 & 00 05 20.14 & -05 42 27.43 &  -8.29~~~   1.64 &   88.19~~~ 0.52 & 25.38~~~   1.05 &-6.56~~~   0.23 &   (1;1;103) \\
         2 & BD+45 4408 &         38 &            & 00 05 41.21 & +45 48 35.34 & 885.60~~~   4.70 & -207.00~~~ 4.75 & 88.60~~~   2.30 &-8.73~    20.10 &   (1;2;13) \\
         3 &       5Cet &        352 &        664 & 00 08 12.10 & -02 26 51.76 &   7.25~~~   1.29 &   -4.87~~~ 0.59 &  3.25~~~   0.96 &-0.43~~~   0.27 &   (1;1;27) \\
         4 &     LN Peg &            &        999 & 00 12 30.31 & +14 33 48.68 & 321.76~~~   1.91 &  -71.31~~~ 0.96 & 24.69~~~   1.20 &-15.30~~~  0.30 &   (1;1;74) \\
         5 &     BD Cet &       1833 &       1792 & 00 22 46.33 & -09 13 51.09 &   3.60~~~   1.27 &  -47.49~~~ 0.83 &  2.40~~~   1.17 & -4.80~~~  0.30 &   (1;1;18) \\
         6 &     13 Cet &       3196 &       2762 & 00 35 14.88 & -03 35 34.25 & 407.68~~~   1.31 &  -36.47~~~ 0.61 & 47.51~~~   1.15 & 10.37~~~  0.40 &   (1;1;122) \\
\hline
\end{tabular}  
}
\end{minipage}
\end{table*}

\begin{table*}
\begin{minipage}{185mm}
\contcaption{}
{\scriptsize
\begin{tabular}{clccrrrrrrr}
\hline 
ID&Name&HD&HIP&$\alpha$(2000)~~~&$\delta$(2000)~~~&$\mu_{\alpha}cos\delta$~~~&$\mu_{\delta}$~~~~~~&  $\pi$~~~~~ &$\gamma$~~~~~~& References \\
        &        &        &       
& (h~m~s)~~~&($^{o}$~~$'$~~$''$)~~~&(mas/yr) &(mas/yr) &(mas)~~~~&(km/s)~~~&            \\
\hline
        61 &  BD+10 828 &      37171 &      26386 & 05 37 04.38 & +11 02 06.02 & 52.25~~~0.80 & -16.18~~~   0.55 & 3.97~~~   0.82 & -120.00~~~   6.66 &   (1;1;26) \\
        62 &     TW Lep &      37847 &      26714 & 05 40 39.72 & -20 17 55.58 & 26.11~~~1.35 & -10.69~~~   1.17 & 5.87~~~   1.19 & 18.00~~~   0.50 &   (1;1;18) \\
        63 &  V1149 Ori &      37824 &      26795 & 05 41 26.79 & +03 46 40.94 & 34.52~~~1.16 & 6.16~~~   0.74 & 6.93~~~   1.13 & 26.90~~~   0.30 &   (1;1;101) \\
        64 &  V1197 Ori &      38099 &      26953 & 05 43 09.32 & -01 36 47.69 & 4.86~~~1.02 & -29.07~~~   0.61 & 4.80~~~   0.95 & 32.06~~~   0.14 &   (1;1;145) \\
        65 &     TZ Col &      39576 &      27727 & 05 52 15.99 & -28 39 24.92 & 5.35~~~0.98 & -19.72~~~   0.78 & 11.41~~~   1.07 & 24.25~~~   3.40 &   (1;1;39) \\
        66 & SAO 234181 &      39937 &      27737 & 05 52 20.20 & -57 09 22.29 & 17.99~~~0.58 & -78.63~~~   0.46 & 7.86~~~   0.50 & 8.00~~~   10.0 &   (1;1;56) \\
\hline
\end{tabular}  
}
\end{minipage}
\end{table*}

\begin{table*}
\begin{minipage}{185mm}
\contcaption{}
{\scriptsize
\begin{tabular}{clccrrrrrrr}
\hline
ID&Name&HD&HIP&$\alpha$(2000)~~~&$\delta$(2000)~~~&$\mu_{\alpha}cos\delta$~~~&$\mu_{\delta}$~~~~~~&  $\pi$~~~~~ &$\gamma$~~~~~~& References \\
        &        &        &            
& (h~m~s)~~~&($^{o}$~~$'$~~$''$)~~~&(mas/yr) &(mas/yr) &(mas)~~~~&(km/s)~~~&            \\
\hline
        121 &     HU Vir &     106225 &     106225 & 12 13 20.69 & -09 04 46.88 & -11.7~~~1.09 & -0.43~~~   0.56 & 8.00~~~   1.25 & -0.66~~~    0.12 &   (1;1;74) \\
       122 &     DK Dra &     106677 &      59796 & 12 15 41.49 & +72 33 04.31 & -9.20~~~0.57 & -25.11~~~   0.50 & 7.24~~~   0.55 & -45.29~~~    0.07 &   (1;1;37) \\
       123 &     AS Dra &     107760 &      60331 & 12 22 11.73 & +73 14 54.54 & -455.39~~~0.82 & 184.34~~~   0.58 & 23.16~~~   0.67 & -98.90~~~    0.10 &   (1;1;122) \\
       124 &     IL Com &     108102 &      60582 & 12 25 02.26 & +25 33 38.36 & -9.94~~~1.12 & -8.97~~~   0.67 & 9.34~~~   1.06 & -0.40~~~    0.50 &   (1;1;113) \\
       125 &     HZ Com &            &            & 12 29 40.92 & +24 31 14.65 & -10.80~~~2.90 & -7.60~~~   2.30 & 16.60~~~   0.83 & -0.56~~~    0.24 &   (6;12;44) \\
       126 &     IM Vir &     111487 &            & 12 49 38.70 & -06 04 44.86 & 26.50~~~1.60 & -57.30~~~   1.60 & 16.67~~~   0.84 & 12.30~~~    2.00 &   (6;9;120) \\
\hline
\end{tabular}  
}
\end{minipage}
\end{table*}

\begin{table*}
\begin{minipage}{185mm}
\contcaption{}
{\scriptsize
\begin{tabular}{clccrrrrrrr}
\hline
ID&Name&HD&HIP&$\alpha$(2000)~~~&$\delta$(2000)~~~&$\mu_{\alpha}cos\delta$~~~&$\mu_{\delta}$~~~~~~&  $\pi$~~~~~ &$\gamma$~~~~~~& References \\
        &        &        &            
& (h~m~s)~~~&($^{o}$~~$'$~~$''$)~~~&(mas/yr) &(mas/yr) &(mas)~~~~&(km/s)~~~&            \\
\hline
       181 &     PW Her &            &      89039 & 18 10 24.11 & +33 24 11.18 & 15.82~~~1.42 & 9.77~~~   1.08 & 4.31~~~    1.29 & -24.40~~~    1.00 &   (1;1;102,136) \\
       182 &     AW Her &     348635 &      90312 & 18 25 38.72 & +18 17 40.25 & 16.31~~~1.50 & 18.37~~~   1.14 & 4.71~~~    1.55 & -45.60~~~    0.15 &   (1;1;108) \\
       183 &     BY Dra &     234677 &      91009 & 18 33 59.45 & +51 43 19.40 & 186.62~~~0.81 & -324.90~~~   0.66 & 60.90~~~    0.73 & -25.43~~~    0.08 &   (1;1;30) \\
       184 &    Omi Dra &     175306 &      92512 & 18 51 12.10 & +59 23 18.07 & 77.56~~~0.56 & 25.43~~~   0.45 & 10.12~~~    0.43 & -19.53~~~    0.31 &   (1;1;52) \\
       185 &     35 Sqr &     175190 &      92845 & 18 55 07.14 & -22 40 16.78 & 110.34~~~1.27 & -30.79~~~   0.99 & 12.07~~~    0.92 & -107.00~~~    4.48 &   (1;1;52) \\
       186 &  V1285 Aql &            &      92871 & 18 55 27.41 & +08 24 09.02 & 92.03~~~2.05 & -69.58~~~   1.42 & 86.27~~~    1.90 & -13.49~~~    0.17 &   (1;1;50) \\
\hline
\end{tabular}  
}
\end{minipage}
\end{table*}
\begin{table*}
\begin{minipage}{185mm}
{\scriptsize
\begin{tabular}{ll}
 & References to Table 1\\
\hline
No & Authors\\ 
\hline
1	&	ESA, 1997, the Hipparcos and Tycho Catalogues, ESA SP-1200\\
2	&	ARI database for nearby stars. Astronomisches Rechen-Institut, Heidelberg (ARICNS)\\
3	&	Bakos, G.A, Sahu, K.C.,\& Nemeth, P., 2002, ApJS, 141, 187\\
4	&	Drake, S.A. et al., 1989, ApJS, 71, 905\\
5	&	Harrington, R.S., \& Dahn, D.C., 1980, AJ, 85, 454\\
6	&	Hog, E. et al., 1998, A\&A, 335, 65\\
\hline
\end{tabular}  
}
\end{minipage}
\end{table*}

\begin{table*}
\begin{minipage}{185mm}
%\contcaption{}
{\scriptsize
\begin{tabular}{ll}
 & References to Table 1\\
\hline
No &	Authors\\ 
\hline
84	&	Griffin, R.F., 1983, JApA, 4, 171\\
85	&	Griffin, R.F., 1985, MNRAS, 212, 663\\
86	&	Griffin, R.F., 1985, Obs., 105, 81\\
87	&	Griffin, R.F., 1988,  Obs., 108, 220\\
88	&	Griffin, R.F., 1988, JApA, 9, 213\\
89	&	Griffin, R.F., 1988, JApA, 9, 75\\
\hline
\end{tabular}  
}
\end{minipage}
\end{table*}

\setcounter{table}{1}
\begin{table*}
\begin{minipage}{185mm}
\caption{Kinematic data of the Chromospherically Active Binaries.}
{\scriptsize
\begin{tabular}{rlrccrrrrrrrrr} 
\hline
ID & Name & HD & {\it l}& {\it b} & X (pc) & Y (pc) & Z (pc) & U (km/s) & V (km/s) & W (km/s) & Code & MG & OC \\
\hline
         1 &     BC Psc &         28 &  93.75& -65.93 &       -1.1 &       16.0 &      -36.0 & -6.97~~    0.39 & 11.07~~    0.60 & 12.31~~    0.34 &            &            &            \\
         2 & BD+45 4408 &         38 & 114.65& -16.32 &       -4.5 &        9.8 &       -3.2 & -34.96~~    4.15 & -31.19~~    8.79 & -16.36~~    2.88 &          1 &        Hya &            \\
         3 &      5 Cet &        352 &  98.34& -63.24 &      -20.1 &      137.1 &     -274.7 & -9.45~~    2.40 & -10.48~~    3.24 & -4.35~~    1.50 &          1 &        Cas &            \\
         4 &     LN Peg &            & 108.98& -47.25 &       -8.9 &       26.0 &      -29.7 & -44.97~~    2.34 & -47.19~~    1.84 & -7.12~~    0.93 &            &            &            \\
         5 &     BD Cet &       1833 & 100.87& -70.86 &      -25.8 &      134.2 &     -393.6 & 37.43~  20.02 & -80.88~  38.75 & -25.19~  14.50 &            &            &            \\
         6 &     13 Cet &       3196 & 112.90& -66.15 &       -3.3 &        7.8 &      -19.3 & -34.81~~    0.81 & -19.90~~    0.60 & -13.49~~    0.38 &          1 &        Hya &            \\
         7 &     BK Psc &            & 118.23& -52.11 &       -9.5 &       17.7 &      -25.9 & -53.22~~    3.29 & -68.92~~    3.70 & -13.98~~    1.35 &            &            &            \\
\hline
\end{tabular}  
}
\end{minipage}
\end{table*}

\begin{table*}
\begin{minipage}{185mm}
\contcaption{}
{\scriptsize
\begin{tabular}{rlrccrrrrrrrrr} 
\hline 
ID & Name & HD & {\it l}& {\it b} & X (pc) & Y (pc) & Z (pc) & U (km/s) & V (km/s) & W (km/s) & Code & MG & OC \\
\hline
        61 &  BD+10 828 &      37171 & 194.33&-11.02 &     -239.6 &      -61.2 &      -48.1 & 119.68~~    6.39 & -19.33~  10.29 & 65.88~~    9.00 &            &            &            \\
        62 &     TW Lep &      37847 & 224.25&-24.32 &     -111.2 &     -108.3 &      -70.2 & -2.66~~    1.47 & -27.19~~    3.36 & 7.58~~    3.20 &            &            &            \\
        63 &  V1149 Ori &      37824 & 201.33&-13.71 &     -130.6 &      -51.0 &      -34.2 & -25.12~~    0.50 & -18.42~~    1.61 & 15.66~~    3.66 &            &            &            \\
        64 &  V1197 Ori &      38099 & 206.44&-15.90 &     -179.4 &      -89.2 &      -57.1 & -10.63~~    2.90 & -36.93~~    4.73 & -17.58~~    1.96 &            &            &            \\
        65 &     TZ Col &      39576 & 233.87&-24.71 &      -46.9 &      -64.3 &      -36.6 & -4.27~~    1.95 & -22.50~~    2.55 & -10.61~~    1.47 &          1 &         LA &            \\
        66 & SAO 234181 &      39937 & 265.48&-30.41 &       -8.6 &     -109.4 &      -64.4 & 50.01~~    3.11 & -14.53~~    8.61 & 2.60~~    5.09 &            &            &            \\
        67 &     SZ Pic &      39917 & 250.03&-28.40 &      -58.6 &     -161.2 &      -92.7 & 61.21~~    6.48 & 8.33~~    2.75 & -0.83~~    2.10 &            &            &            \\
\hline
\end{tabular}  
}
\end{minipage}
\end{table*}

\begin{table*}
\begin{minipage}{185mm}
\contcaption{}
{\scriptsize
\begin{tabular}{rlrccrrrrrrrrr} 
\hline 
ID & Name & HD & {\it l}& {\it b} & X (pc) & Y (pc) & Z (pc) & U (km/s) & V (km/s) & W (km/s) & Code & MG & OC \\
\hline
       121 &     HU Vir &     106225 & 287.32  & 52.63 &       22.6 &      -72.4 &       99.3 & -4.05~~    1.09 & -3.11~~    0.66 & -1.70~~    0.30 &          1 &        Cas &            \\
       122 &     DK Dra &     106677 & 126.66 &  44.32 &      -59.0 &       79.3 &       96.5 & 16.14~~    0.34 & -39.67~~    1.09 & -20.93~~    0.85 &            &            &            \\
       123 &     AS Dra &     107760 & 125.84  & 43.71 &      -18.3 &       25.3 &       29.8 & -49.66~~    2.63 & -79.29~~    0.63 & -105.84~~    1.09 &            &            &            \\
       124 &     IL Com &     108102 & 226.29  & 83.88 &       -7.9 &       -8.3 &      106.5 & -1.82~~    0.57 & -6.39~~    0.84 & -1.05~~    0.51 &          1 &        Cas &    Com Ber \\
       125 &     HZ Com &            & 239.77  & 84.45 &       -2.9 &       -5.0 &       60.0 & -1.38~~    0.79 & -3.37~~    0.72 & -0.92~~    0.25 &          1 &        Cas &    Com Ber \\
       126 &     IM Vir &     111487 & 302.12  & 56.79 &       17.5 &      -27.8 &       50.2 & 18.10~~    1.00 & -13.11~~    1.08 & 1.42~~    1.75 &            &            &            \\
       127 &     IN Com &     112313 & 339.90  & 88.46 &       17.9 &       -6.6 &      709.0 & -69.20~  18.91 & -45.16~  12.59 & -15.17~~    0.42 &            &            &            \\
\hline
\end{tabular}  
}
\end{minipage}
\end{table*}

\begin{table*}
\begin{minipage}{185mm}
\contcaption{}
{\scriptsize
\begin{tabular}{rlrccrrrrrrrrr} 
\hline 
ID & Name & HD & {\it l}& {\it b} & X (pc) & Y (pc) & Z (pc) & U (km/s) & V (km/s) & W (km/s) & Code & MG & OC \\
\hline
       181 &     PW Her &            & 60.13  & 22.54 &      106.7 &      185.8 &       88.9 & -27.40~~    3.53 & -7.63~~    3.83 & -21.44~~    3.90 &            &            &            \\
       182 &     AW Her &     348635 & 46.68  & 13.77 &      141.5 &      150.0 &       50.5 & -50.46~~    5.33 & -15.09~~    5.82 & -17.95~~    2.73 &          1 &        Hya &            \\
       183 &     BY Dra &     234677 & 80.56  & 23.57 &        2.5 &       14.8 &        6.6 & 17.68~~    0.27 & -18.38~~    0.10 & -28.86~~    0.23 &            &            &            \\
       184 &    Omi Dra &     175306 & 89.31  & 23.14 &        1.1 &       90.9 &       38.8 & -23.20~~    0.90 & -5.07~~    0.63 & -37.25~~    1.29 &          1 &         IC &            \\
       185 &     35 Sqr &     175190 & 12.91 & -10.89 &       79.3 &       18.2 &      -15.7 & -112.73~~    4.36 & -18.35~~    1.13 & -23.37~~    3.46 &            &            &            \\
       186 &  V1285 Aql &            & 40.87  &   2.93 &       8.8 &        7.6 &        0.6 & -9.43~~    0.14 & -9.47~~    0.13 & -6.92~~    0.17 &          1 &        Cas &            \\
       187 &   V775 Her &     175742 & 54.58   &  9.58 &      12.3 &       17.2 &        3.6 & 24.26~~    0.42 & -0.16~~    0.20 & -22.34~~    0.54 &            &            &            \\
\hline
\end{tabular} 
}
\end{minipage}
\end{table*} 
\setcounter{table}{3}
\begin{table*}
\begin{minipage}{185mm}
\caption{Physical parameters of CABs.}
{\scriptsize
\begin{tabular}{clllcrcrccr}
\hline
ID& Name & HD & Spectral type & SB & $P$(days) &  $e$ & $M_{h}/M_{c}$ & $M_{c}$ & $f(M)$ & $R_{h}/R_{c}$ \\
\hline

         1 &     BC Psc &         28 &     K0IIIb &        SB1 &      72.93 &      0.270 &            &  0.42-2.00 &     0.0300 &      /14.7 \\
         2 & BD+45 4408 &         38 &        dK6 &        SB1 &            &            &            &            &            &            \\
         3 &      5 Cet &        352 &      K2III &        SB1 &      96.40 &      0.040 &       0.79 &       1.40 &     0.1400 &     /39-42 \\
         4 &     LN Peg &            &    G8V+K5V & SB1 in SB3 &       1.84 &      0.000 &            &            &     0.4000 &      0.76/ \\
         5 &     BD Cet &       1833 &    K1III+F &        SB1 &      35.10 &      0.040 &            &            &     0.1100 &      /14.0 \\
         6 &     13 Cet &       3196 &    F8V+G4V &        SB1 &       2.08 &      0.000 &       1.83 &       0.24 &     0.0189 &      /1.47 \\
         7 &     BK Psc &            &    K5V+M4V &        SB1 &       2.17 &      0.003 &       1.81 &       0.37 &            &  0.72/0.45 \\
\hline
\end{tabular}  
}
\end{minipage}
\end{table*}

\begin{table*}
\begin{minipage}{185mm}
\contcaption{}
{\scriptsize
\begin{tabular}{clllcrcrccr}
\hline
ID& Name & HD & Spectral type & SB & $P$(days) &  $e$ & $M_{h}/M_{c}$ & $M_{c}$ & $f(M)$ & $R_{h}/R_{c}$ \\
\hline
        61 &  BD+10 828 &      37171 &      K4III &        SB1 &            &            &            &            &            &            \\
        62 &     TW Lep &      37847 & F6IV+K2III &        SB2 &      28.34 &      0.050 &       0.95 &       1.02 &            &      /11.5 \\
        63 &  V1149 Ori &      37824 &  K1III+F5V &        SB1 &      53.58 &      0.000 &            &            &     0.0900 &      /13.4 \\
        64 &  V1197 Ori &      38099 &      K4III &        SB1 &     143.04 &      0.000 &       0.79 &       0.90 &     0.0112 &      /51.3 \\
        65 &     TZ col &      39576 &        G1V &        SB1 &            &            &            &            &            &     /$>$1.07 \\
        66 & SAO 234181 &      39937 &       F7IV &            &            &            &            &            &            &            \\
        67 &     SZ Pic &      39917 & K0IV-III+G3IV+III&  SB2 &       4.96 &            &            &            &            &            \\
\hline
\end{tabular}  
}
\end{minipage}
\end{table*}

\begin{table*}
\begin{minipage}{185mm}
\contcaption{}
{\scriptsize
\begin{tabular}{clllcrcrccr}
\hline
ID& Name & HD & Spectral type & SB & $P$(days) &  $e$ & $M_{h}/M_{c}$ & $M_{c}$ & $f(M)$ & $R_{h}/R_{c}$ \\
\hline
       121 &     HU Vir &     106225 &       K0IV & SB1 in SB3 &      10.39 &      0.009 &            &            &     0.1027 &      /3.40 \\
       122 &     DK Dra &     106677 & K1III+K1III&        SB2 &      64.47 &      0.000 &       1.02 &       1.72 &            &  14.0/14.0 \\
       123 &     AS Dra &     107760 &    G4V+G9V &        SB2 &       5.41 &      0.000 &       1.13 &       0.71 &            &            \\
       124 &     IL Com &     108102 &    F8V+F8V &        SB2 &       0.96 &      0.000 &       1.04 &       0.82 &            &  1.10/1.10 \\
       125 &     HZ Com &            &     G9+K4V &        SB2 &       3.56 &      0.013 &       1.05 &       0.67 &            & $>$0.85/$>$1.1 \\
       126 &     IM Vir &     111487 &        G5V &        SB1 &       1.31 &      0.000 &            &            &     0.1060 &            \\
       127 &     IN Com &     112313 &   G5III-IV & SB2 in SB3 &       1.99 &      0.000 &       3.64 &   $>$0.004 &            &  1.76/0.58 \\
\hline
\end{tabular}  
}
\end{minipage}
\end{table*}

\begin{table*}
\begin{minipage}{185mm}
\contcaption{}
{\scriptsize
\begin{tabular}{clllcrcrccr}
\hline
ID& Name & HD & Spectral type & SB & $P$(days) &  $e$ & $M_{h}/M_{c}$ & $M_{c}$ & $f(M)$ & $R_{h}/R_{c}$ \\
\hline
       181 &     PW Her &            & K0IV+F8-G2 &        SB2 &       2.88 &      0.000 &       0.77 &       1.50 &            &    1.4/3.8 \\
       182 &     AW Her &     348635 &  G2IV+K2IV &        SB2 &       8.80 &      0.000 &       0.91 &       1.33 &            &    2.4/3.2 \\
       183 &     BY Dra &     234677 & K6Ve + K7V &        SB2 &       5.98 &      0.307 &       1.12 &       0.44 &            & 1.20-1.40/ \\
       184 &    Omi Dra &     175306 &      G9III &        SB1 &     138.42 &      0.114 &            &            &     0.1830 &  /$>$37.00 \\
       185 &     35 Sqr &     175190 &      K3III &        SB1 &            &            &            &            &            &            \\
       186 &  V1285 Aql &            &   dM2e+dMe &        SB2 &      10.32 &      0.200 &       1.07 &       0.30 &            &  0.44/0.44 \\
       187 &   V775 Her &     175742 &  K0Ve+dM3e &        SB1 &       2.88 &      0.003 &            &            &     0.0362 &       /0.9 \\
\hline
\end{tabular}  
}
\end{minipage}
\end{table*}
\end{document}